\documentstyle[12pt]{article}

\begin{document}

\begin{flushright}
Preprint: imsc-94/43 \\
October 1994
\end{flushright}

\noindent {\large \bf On a nonstandard two-parametric quantum
algebra  \\
and its connections with $U_{p,q}(gl(2))$ and
$U_{p,q}(gl(1|1))$}

\bigskip

\noindent R. Chakrabarti$^1$, R. Jagannathan$^2$
\footnote{E-mail:\ jagan@imsc.ernet.in}

\medskip

\noindent $^1${\small Department of Theoretical Physics,
University of Madras,
Guindy Campus, \\
Madras-600025, India}

\vspace{.25cm}

\noindent $^2${\small The Institute of Mathematical Sciences,
C I T Campus, Tharamani, \\
Madras-600113, India}

\bigskip

\baselineskip18pt

\noindent {\bf Abstract.}  A quantum algebra
$U_{p,q}(\zeta ,H,X_\pm )$ associated with a nonstandard
$R$-matrix with two deformation parameters$(p,q)$ is studied and,
in particular, its universal ${\cal R}$-matrix is derived using
Reshetikhin's method.  Explicit construction of the
$(p,q)$-dependent nonstandard $R$-matrix is obtained through a
coloured generalized boson realization of the universal
${\cal R}$-matrix of the standard $U_{p,q}(gl(2))$ corresponding
to a nongeneric case.  General finite dimensional coloured
representation of the universal ${\cal R}$-matrix of
$U_{p,q}(gl(2))$ is also derived.  This representation, in
nongeneric cases, becomes a source for various $(p,q)$-dependent
nonstandard $R$-matrices.  Superization of
$U_{p,q}(\zeta , H,X_\pm )$ leads to the super-Hopf algebra
$U_{p,q}(gl(1|1))$.  A contraction procedure then yields a
$(p,q)$-deformed super-Heisenberg algebra $U_{p,q}(sh(1))$ and
its universal ${\cal R}$-matrix.

\vspace{1cm}

\noindent{\bf 1. Introduction}

\medskip

\noindent The single parameter quantization of the universal
enveloping algebra of a simple Lie algebra is well-known [1].
The Yang-Baxter equation (YBE), however, also admits nonstandard
solutions [2-4] characterizing quasitriangular Hopf algebras, which
are not deformations of classical algebras.  The nonstandard quantum
algebra $U_q(\zeta , H,X_\pm )$ associated with the Alexander-Conway
solution of the YBE has been studied [2,3] and the relevant universal
${\cal R}$-matrix has been obtained [4].  Using transmutation theory
[5] it has been argued [4] that the superized
$U_q(\zeta ,H,X_\pm)$ coincides with a super-Hopf algebra
$U_q(gl(1|1))$.  Moreover, using a general $q$-boson [6]
realization of the Hopf algebra $U_q(gl(2))$, it has been observed
[7] that the nonstandard $R$-matrix of $U_q(\zeta ,H,X_\pm )$ may
be obtained from the universal ${\cal R}$-matrix of $U_q(gl(2))$
in a nongeneric limit.

In another development, the constructions and
representations of quantum algebras with multiple deformation
parameters have been studied extensively [8-15].  For quasitriangular
Hopf algebras, Reshetikhin [9] has developed a general formalism to
introduce multiple deformation parameters.  Following [9] the
universal ${\cal R}$-matrix of the quantum algebra $U_{p,q}(gl(2))$
with two independent parameters $(p,q)$ has been obtained [15].  An
identical procedure may be adopted to construct the universal
${\cal R}$-matrix of the super-Hopf algebra $U_{p,q}(gl(1|1))$; this
verifies the known result obtained by direct computation [13].  Here,
we study a $(p,q)$-generalization, $U_{p,q}(\zeta ,H,X_\pm )$, of the
Alexander-Conway algebra $U_q(\zeta ,H,X_\pm )$.

Following the
prescription in [9] the universal ${\cal R}$-matrix for the
quasitriangular Hopf algebra $U_{p,q}(\zeta ,H,X_\pm )$ is obtained
in section 2.  The nonstandard Hopf algebra $U_{p,q}\left( \zeta ,H,
X_\pm \right)$ has been previously considered in [16].  These authors,
however, have not discussed the universal ${\cal R}$-matrix for
$U_{p,q}\left(\zeta ,H,X_\pm \right)$.  Parallel to its single
deformation parameter analogue,
$U_{p,q}(\zeta ,H,X_\pm )$ exhibits close kinships with
$U_{p,q}(gl(2))$ and $U_{p,q}(gl(1|1))$.  In particular, the
nonstandard $R$-martrix of $U_{p,q}\left(\zeta, H, X_\pm \right)$
is obtained in section 3 using a coloured generalized boson
representation of the universal ${\cal R}$-matrix of
$U_{p,q}(gl(2))$ in a nongeneric limit.  A general recipe for
realizing the finite dimensional nonstandard two-parametric coloured
$R$-matrices associated with nongeneric representations of
$U_{p,q}(gl(2))$ is also presented.  In section 4, a map connecting
the two-parametric quantum algebras
$U_{p,q}\left( \zeta, H, X_\pm \right)$ and $U_{p,q}(gl(1|1))$ via
the superizatiion procedure is described.  A contraction procedure
is used in section 5 to extract a $(p,q)$-deformed quasitriangular
super-Heisenberg algebra $U_{p,q}(sh(1))$.  We conclude in section 6.

\bigskip

\noindent {\bf 2. Quasitriangular Hopf algebra
$U_{p,q}\left( \zeta , H, X_\pm \right)$}

\medskip

\noindent We study the Hopf algebra associated with the two-parametric
nonstandard solution [16] of the YBE

\renewcommand{\theequation}{2.{\arabic{equation}}}
\setcounter{equation}{0}

\begin{equation}
R\,=\,\left(
\begin{array}{cccc}
Q^{-1} & 0 & 0 & 0 \\
0 & \lambda ^{-1} & 0 & 0 \\
0 & \sigma & \lambda & 0 \\
0 & 0 & 0 & -Q
\end{array}
\right)\,, \qquad
\sigma \,=\,Q^{-1}-Q\,.
\end{equation}

\noindent The defining relation of the quantum inverse
scattering method [17]

\begin{equation}
R(T \otimes 1\!{\rm l})(1\!{\rm l} \otimes T)\,=\,(1\!{\rm l}
 \otimes T)(T \otimes 1\!{\rm l})R\,,
\end{equation}

\noindent with the $R$-matrix as given by (2.1), describes a
transfer matrix

\begin{equation}
T\,=\,\left(
\begin{array}{cc}
a & b \\
c & d
\end{array}
\right)\,,
\end{equation}

\noindent whose elements obey the braiding relations

\begin{eqnarray}
ab & = & p^{-1}ba\,, \quad ac\,=\,q^{-1}ca\,,
\quad db\,=\,-p^{-1}bd\,, \quad dc\,=\,-q^{-1}cd\,,  \nonumber \\
p^{-1}bc & = & q^{-1}cb\,, \quad ad-da\,=\,(p^{-1}-q)bc\,, \quad
b^2\,=\,0\,, \quad c^2\,=\,0\,,
\end{eqnarray}

\noindent where

\begin{equation}
p\,=\,\lambda Q\,, \qquad q\,=\,\lambda ^{-1}Q\,.
\end{equation}

A conjugate $R$-matrix

\begin{equation}
\tilde{R}\,=\,\left( R^{(+)} \right)^{-1}\,=\,\left(
\begin{array}{cccc}
Q & 0 & 0 & 0 \\
0 & \lambda ^{-1} & -\sigma & 0 \\
0 & 0 & \lambda & 0 \\
0 & 0 & 0 & -Q^{-1}
\end{array}
\right)
\end{equation}

\noindent also fits (2.2) with the elements of $T$ obeying (2.4).  The
matrices $R^{\left( \pm \right)}$ are defined by

\begin{equation}
R^{(+)}\,=\,PRP\,,  \qquad
R^{(-)}\,=\,R^{-1}\,,
\end{equation}

\noindent where $P$ is the permutation matrix, given by

\begin{equation}
P\,=\,\left(
\begin{array}{cccc}
1 & 0 & 0 & 0 \\
0 & 0 & 1 & 0 \\
0 & 1 & 0 & 0 \\
0 & 0 & 0 & 1
\end{array}
\right)\,.
\end{equation}

If $a$ and $d$ are invertable, the elements $\left\{
a,b,c,d,a^{-1},d^{-1}\right\}$ generate a Hopf algebra
$A_{p,q}(R)$ (or $Fun_{p,q}(R)$) whose coalgebraic structure readily
follows.  The coproduct$(\Delta )$, counit$(\epsilon )$ and
antipode$(S)$  maps are, respectively, given by

\begin{eqnarray}
\Delta (T) & = & T \dot{\otimes} T\,, \nonumber \\
\Delta \left(a^{-1} \right) & = & a^{-1} \otimes a^{-1} - a^{-1}b
a^{-1} \otimes a^{-1}ca^{-1}\,, \nonumber \\
\Delta \left(d^{-1} \right) & = & d^{-1} \otimes d^{-1} - d^{-1}cd^{-1}
\otimes d^{-1}bd^{-1}\,,
\end{eqnarray}

\begin{equation}
\epsilon (T)\,=\,1\!{\rm l}\,,
\end{equation}

\begin{equation}
S(T)\,=\,T^{-1}\,, \qquad S\left( a^{-1} \right)\,=\,a-bd^{-1}c\,,
\qquad S\left( d^{-1} \right)\,=\,d-ca^{-1}b\,,
\end{equation}

\noindent where

\begin{equation}
T^{-1}\,=\,\left( \begin{array}{cc}
                   a^{-1}+a^{-1}bd^{-1}ca^{-1} & -a^{-1}bd^{-1} \\
                   -d^{-1}ca^{-1}  &  d^{-1}+d^{-1}ca^{-1}bd^{-1}
                   \end{array} \right)
\end{equation}

\noindent and $\dot{\otimes}$ denotes the tensor product coupled with
usual matrix multiplication.  In the Hopf algebra $A_{p,q}(R)$ an
invertable `group-like' element $D$ exists:

\begin{equation}
D\,=\,ad^{-1}-bd^{-1}cd^{-1}\,,
\qquad D^{-1}\,=\,da^{-1}-ba^{-1}ca^{-1}\,.
\end{equation}

\noindent Using (2.4), the commutation relations for $D$ follow:

\begin{equation}
[D,a\,]\,=\,0\,, \qquad [D,d\,]\,=\,0\,, \qquad
\{D,b\}\,=\,0\,, \qquad \{D,c\}\,=\,0\,.
\end{equation}

\noindent The induced coalgebra maps for $D$ are obtained from the
relations (2.9-2.12):

\begin{equation}
\Delta (D)\,=\,D \otimes D\,, \qquad \epsilon (D)\,=\,1\,, \qquad
S(D)\,=\,D^{-1}\,.
\end{equation}

Using the FRT-approach [17], the commutation relations for the
generators $\left( \zeta ,H,X_\pm \right)$ of the Hopf algebra
$U_{p,q}\left( \zeta ,H,X_\pm \right)$, dual to the algebra
$A_{p,q}(R)$, are obtained from the relations

\begin{equation}
R^{(+)}\left( L^{\left( \varepsilon _1 \right)} \otimes 1\!{\rm l}
\right) \left( 1\!{\rm l} \otimes L^{\left( \varepsilon _2 \right)}
\right)\,=\,\left( 1\!{\rm l} \otimes L^{\left( \varepsilon _2 \right)}
\right) \left( L^{\left( \varepsilon _1 \right)} \otimes 1\!{\rm l}
\right)R^{(+)}\,,
\end{equation}

\noindent where  $\left( \varepsilon _1,\varepsilon _2 \right)$ $=$
$(+,+)$, $(-,-)$, $(+,-)$ and

\begin{equation}
L^{(+)}\,=\,\left( \begin{array}{cc}
                   p^{-H}q^{-\zeta } & \sigma p^{-H-\frac{1}{2}}X_- \\
                           0     & gp^{-H}q^\zeta
                   \end{array} \right)\,,  \quad
L^{(-)}\,=\,\left( \begin{array}{cc}
                   q^Hp^\zeta     &   0  \\
                   -\sigma gq^{H-\frac{1}{2}}X_+ & gq^Hp^{-\zeta }
                   \end{array} \right)\,,
\end{equation}

\noindent with $g\,=\,(-1)^{\zeta -H}$.  The corresponding commutation
relations read

\begin{eqnarray}
X_\pm ^2 & = & 0\,, \quad \left[H,X_\pm \right]\,=\,\pm X_\pm \,,
\quad
\left\{ X_+,X_- \right\}\,=\,[2\zeta ]\,, \quad \nonumber \\
\left[ \zeta , X \right] & = & 0 \qquad
\forall X\,=\,H,X_\pm \,,
\end{eqnarray}

\noindent where

\begin{equation}
[X]\,=\,\frac{Q^X - Q^{-X}}{Q-Q^{-1}}\,.
\end{equation}

\noindent The comultiplication maps for the generators are

\begin{eqnarray}
\Delta \left( X_\pm \right) & = & X_\pm \otimes g^{\mp 1}Q^\zeta
\lambda ^{\pm \zeta } + Q^{-\zeta }\lambda ^{\mp \zeta } \otimes
X_\pm \,, \nonumber \\
\Delta (H) & = & H \otimes 1\!{\rm l} + 1\!{\rm l} \otimes H\,, \qquad
\Delta (\zeta )\,=\,\zeta \otimes 1\!{\rm l} + 1\!{\rm l} \otimes
\zeta \,,
\end{eqnarray}

\noindent which follow from the relations

\begin{equation}
\Delta \left( L^{\left(\pm \right)} \right)\,=\,L^{\left( \pm \right)}
\dot{\otimes } L^{\left( \pm \right)}\,.
\end{equation}

\noindent The counit and the antipode maps are given by

\begin{equation}
\epsilon (X)\,=\,0\,, \qquad \forall X\,=\,\zeta ,H,X_\pm \,,
\end{equation}

\begin{equation}
S(\zeta )\,=\,-\zeta \,, \qquad S(H)\,=\,-H\,,  \qquad
S\left( X_\pm \right)\,=\,g^{\pm 1}X_\pm \,.
\end{equation}

In spite of the appearance of anticommutator in (2.18), the
Hopf algebra $U_{p,q}\left( \zeta ,H,X_\pm \right)$ is bosonic as it
follows the direct product rule

\begin{equation}
(A \otimes B)(C \otimes D)\,=\,AC \otimes BD\,,  \qquad
\forall A,B,C,D \in U_{p,q}\left( \zeta , H,X_\pm \right)\,.
\end{equation}

\noindent For the finite dimensional faithful representation [3] of
the algebra (2.18)

\begin{eqnarray}
X_+ & = & \left( \begin{array}{cc}
               0 & [z] \\
               0 & 0
               \end{array} \right)\,, \qquad
X_- = \left( \begin{array}{cc}
               0 & 0 \\
               1 & 0
               \end{array} \right)\,, \qquad
H = \frac{1}{2}\left( \begin{array}{cc}
                        1 & 0 \\
                        0 & -1
                        \end{array} \right)\,,
\nonumber \\
\zeta & = & \frac{z}{2}\left( \begin{array}{cc}
                             1 & 0 \\
                             0 & 1
                             \end{array} \right)\,,
\end{eqnarray}

\noindent with $z \in {\rm C}\!\!\!{\rm C}$\,\,,
the duality relations between the Hopf algebras $A_{p,q}(R)$ and
$U_{p,q}\left( \zeta ,H,X_\pm \right)$ assume the form

\begin{equation}
\left\langle L^{\left( \pm \right)}\,,\,T \right\rangle \,=\,R^{\left(
\pm \right)}\,,
\end{equation}

\noindent for the choice $z\,=\,1$.

We now discuss the quasitriangular character of
$U_{p,q}\left( \zeta ,H,X_\pm \right)$ containing a group-like
element $g$, that is, now and henceforth, assumed to satisfy
$g^2$ $=$ $1$. For a quasitriangular Hopf algebra ${\cal U}$, the
universal ${\cal R}$-matrix
$\left( \in {\cal U} \otimes {\cal U} \right)$ satisfies
the relations

\begin{eqnarray}
\tau \circ \Delta (X) & = & {\cal R} \Delta (X) {\cal R}^{-1}\,, \qquad
\tau \circ (X \otimes Y)\,=\,Y \otimes X\,, \quad
\forall\,\,X,Y \in {\cal U}\,, \nonumber \\
(\Delta \otimes {\rm id}){\cal R} & = & {\cal R}_{13}{\cal R}_{23}\,,
\qquad ({\rm id} \otimes \Delta ){\cal R}\,=\,{\cal R}_{13}
{\cal R}_{12}\,, \nonumber \\
(\epsilon \otimes {\rm id}) & = & ({\rm id} \otimes \epsilon )
{\cal R}\,=\,1\!{\rm l}\,,  \nonumber \\
(S \otimes {\rm id}){\cal R} & = & {\cal R}^{-1}\,,  \qquad
({\rm id} \otimes S){\cal R}^{-1}\,=\,{\cal R}\,,
\end{eqnarray}

\noindent where the subscripts in ${\cal R}_{ij}$ indicate the embedding
of ${\cal R}$ in ${\cal U}^{\otimes 3}$.  The explicit expression for
the universal ${\cal R}$-matrix of $U_Q\left( \zeta ,H,X_\pm \right)$
is [4]

\begin{equation}
{\cal R}_{\lambda =1} = (-1)^{(\zeta - H) \otimes (\zeta - H)}
Q^{2(\zeta \otimes H + H \otimes \zeta )} \left( 1\!{\rm l} \otimes
1\!{\rm l} + \sigma Q^\zeta X_+ \otimes gQ^{-\zeta }X_- \right)\,.
\end{equation}

\noindent Employing the Reshetikhin procedure [9] we obtain the universal
${\cal R}$-matrix for $U_{p,q}\left( \zeta
,H,X_\pm \right)$\,.   To this end, we note that the coproduct relations
(2.20) for the generators of $U_{p,q}\left( \zeta ,H,X_\pm \right)$ and the
corresponding relations for the generators of
$U_Q\left( \zeta ,H,X_\pm \right)$, obtained in
the limit $\lambda $ $=$ $1$, are related by a similarity transformation

\begin{equation}
\Delta (X)\,=\,F\left( \Delta _{\lambda =1}(X) \right)F^{-1}\,, \qquad
\forall \,\,X\,=\,\zeta ,H,X_\pm \,,
\end{equation}

\noindent where

\begin{equation}
F\,=\,\lambda ^{(H \otimes \zeta - \zeta \otimes H)}\,.
\end{equation}

\noindent Then, the procedure in [9] yields the universal
${\cal R}$-matrix for $U_{p,q}\left( \zeta ,H, X_\pm \right)$

\begin{equation}
{\cal R}\,=\,F^{-1}{\cal R}_{\lambda =1}F^{-1}
\end{equation}

\noindent which, by construction, satisfies the required relations (2.27).
Explicitly, the universal ${\cal R}$-matrix of $U_{p,q}\left( \zeta ,H,
X_\pm \right)$ reads

\begin{eqnarray}
{\cal R} & = & (-1)^{(\zeta -H) \otimes (\zeta -H)} Q^{2(\zeta \otimes H
+ H \otimes \zeta )} \lambda ^{2(\zeta \otimes H - H \otimes \zeta )}
\nonumber \\
&  & \ \ \times  \left( 1\!{\rm l} \otimes 1\!{\rm l} +
\sigma Q^\zeta \lambda ^\zeta
X_+ \otimes g Q^{-\zeta } \lambda ^\zeta X_- \right)\,.
\end{eqnarray}

\noindent For the representation (2.25) with the choice $z = 1$, ${\cal R}$
in (2.32) reduces to the matrix $\tilde{R}$ in (2.6), thus expressing an
aspect of the duality of $A_{p,q}(R)$ and
$U_{p,q}\left( \zeta ,H,X_\pm \right)$\,.

\bigskip

\noindent {\bf 3. Coloured $R$-matrices associated with nongeneric
representations of $U_{p,q}(gl(2))$}

\medskip

\noindent After completing the above construction of the dual Hopf algebras
corresponding to the nonstandard solution (2.1) of the YBE, we now relate
the $R$-matrix (2.1) to the representations of $U_{p,q}(gl(2))$ for
nongeneric values of $Q$, namely, roots of unity.  Here, we closely
follow the treatment of $U_q(gl(2))$ by Ge et al. [7].  These authors
have constructed a general parameter dependent finite dimensional
$q$-boson realization of the universal ${\cal R}$-matrix of $U_q(gl(2))$;
thereby generating the nonstandard $R$-matrices at $q$, a root of unity.
When different parameters are chosen for different components of the
representation module, the resulting finite dimensional $R$-matrices
are said to be coloured and they obey a coloured YBE [7].

In the standard Hopf algebraic structure of $U_{p,q}(gl(2))$
(see, e.g., [15]) the commutation relations between the generators are

\renewcommand{\theequation}{3.{\arabic{equation}}}
\setcounter{equation}{0}

\begin{equation}
\left[ J_0 , J_\pm \right] = \pm J_\pm \,, \quad \left[ J_+ , J_- \right] =
\left[ 2J_0 \right]\,, \quad [ Z , X ] = 0\,, \ \
\forall \,\,X \in \left( J_0 , J_\pm \right)\,,
\end{equation}

\noindent and the comultiplication rules are

\begin{eqnarray}
\Delta \left( X_\pm \right) & = & J_\pm \otimes Q^{J_0} \lambda ^{\pm Z}
+ Q^{-J_0} \lambda ^{\mp Z} \otimes J_\pm \,, \nonumber \\
\Delta \left( J_0 \right) & = & J_0 \otimes 1\!{\rm l} + 1\!{\rm l}
\otimes J_0\,, \qquad \Delta (Z)\,=\,Z \otimes 1\!{\rm l} + 1\!{\rm l}
\otimes Z\,.
\end{eqnarray}

\noindent The universal ${\cal R}$-matrix of $U_{p,q}(gl(2))$ reads [15]

\begin{eqnarray}
\bar{\cal R} & = & Q^{2\left( J_0 \otimes J_0 \right)} \lambda ^{2\left(
Z \otimes J_0 - J_0 \otimes Z \right)} \nonumber \\
&  & \ \ \times \sum _{n=0}^{\infty}\,\frac{\left( 1-Q^{-2} \right)^n}
{[n]!}\,Q^{\frac{1}{2}n(n-1)} \left( Q^{J_0} \lambda ^Z J_+ \otimes
Q^{-J_0} \lambda ^Z J_- \right)^n\,,
\end{eqnarray}

\noindent where $[n]! = [n][n-1]\ldots [2][1]$\,.

The generators of the
deformed boson algebra [6] satisfy

\begin{equation}
\left[ N , a_\pm \right]\,=\,\pm a_\pm \,, \qquad
a_+a_-\,=\,[N]\,,  \qquad
a_-a_+\,=\,[N+1]\,,
\end{equation}

\noindent and the map [7]

\begin{equation}
J_+\,=\,a_+[\omega -N]\,,  \qquad
J_-\,=\,a_-\,,  \qquad J_0\,=\,N-\frac{\omega}{2}\,,
\quad \omega \in {\rm C}\!\!\!{\rm C}\,,
\end{equation}

\noindent provides an infinite dimensional representation of the algebra
(3.1) in the Fock space ${\cal F}_z \left\{ |m\,z\rangle = a_+^{\,m}|0\,z
\rangle \right.$ $ \left. \mid a_-|0\,z\rangle = 0, N|0\,z\rangle = 0,
m \in {\rm Z}\!\!\!{\rm Z}^+, z \in {\rm C}\!\!\!{\rm C} \right\}$:

\begin{eqnarray}
J_+|m\,z\rangle & = & [\omega -m]|m+1\,z\rangle \,, \qquad
J_-|m\,z\rangle \,=\,[m]|m-1\,z\rangle \,, \nonumber \\
J_0|m\,z\rangle & = & \left( m - \frac{\omega}{2} \right)|m\,z\rangle \,,
\qquad Z|m\,z\rangle \,=\,z|m\,z\rangle \,.
\end{eqnarray}

\noindent The parameter $\omega$ is called colour and provides the key
to obtain the nonstandard $R$-matrix (2.1) starting from the universal
${\cal R}$-matrix of $U_{p,q}(gl(2))$, namely $\bar{\cal R}$ in (3.3).

For the nongeneric
cases $\left\{ Q^n = \pm 1, n \in {\rm Z}\!\!\!{\rm Z}^+ \right\}$ the
identity $[\alpha n] = 0$ holds for $\alpha $ $\in$ ${\rm Z}\!\!\!
{\rm Z}^+$ suggesting the existence of an extremal vector $\{ |\alpha
n\,z\rangle \mid $ $J_-|\alpha n\,z\rangle $ $= 0 \}$.  The corresponding
invariant subspace that renders the representation (3.6) reducible is
${\cal V}_{\alpha \,z}\left\{ |\alpha n+m\,z
\rangle \mid m \in {\rm Z}\!\!\!\!{\rm Z}^+ \right\}$.
  Then, on the quotient space $V_{J\,z}$ $=$
${\cal F}_z/{\cal V}_{\alpha \,z}
\left\{ |J\,M\,z\rangle (=|m\,z \rangle ) \mid m = 0,1,
\ldots ,\right. $ $(\alpha n-1)=2J,$ $\left. M=m-J \right\}$ a finite
dimensional representation for algebra (3.1) holds:

\begin{eqnarray}
J_+|J\,M\,z \rangle & = & [\omega -J-M]\,|J\,(M+1)\,z\rangle
\theta (J-1-M)\,,  \nonumber \\
J_-|J\,M\,z \rangle & = & [J+M]\,|J\,(M-1)\,z\rangle \,, \nonumber \\
J_0|J\,M\,z \rangle & = & \left( J+M-\frac{\omega}{2}\right)
|J\,M\,z\rangle \,,\nonumber \\
Z|J\,M\,z \rangle & = & z|J\,M\,z\rangle \,,
\end{eqnarray}

\noindent where $\theta (x) = 1\,(0)$ for $x \geq 0\,(<0)$\,.  For
$\alpha$ $=$ $1$, the representation (3.7) is irreducible and for $\alpha$
$\geq$ $2$, it is indecomposable [7].  Now, using the representation (3.7),
$\bar{\cal R}$ in (3.3) may be written in the matrix form

\begin{equation}
\bar{\cal R}\left| J_1\,M_1\,z_1 \right\rangle \otimes \left|J_2\,M_2\,z
_2\right\rangle =\!\!\sum _{M_1^{\,\prime}\,M_2^{\,\prime}}
\left( \bar{\cal R}^{J_1\,z_1\,\omega _1\,,\,J_2\,z_2\,\omega _2} \right)
_{M_1\,M_2}^{M_1^{\,\prime}\,M_2^{\,\prime}} \left| J_1\,M_1^{\,\prime}\,z
_1 \right\rangle \otimes \left|J_2\,M_2^{\,\prime}\,z_2\right\rangle
\end{equation}

\noindent where different representations and colour parameters are
chosen in the two sectors of the tensor product space.  Explicitly we
have

\begin{eqnarray}
&    &
\left( \bar{\cal R}^{J_1\,z_1\,\omega _1\,,\,J_2\,z_2\,\omega _2}\right)
_{M_1\,M_2}^{M_1^{\,\prime}\,M_2^{\,\prime}} \nonumber \\
&  &  \quad =  Q^{2\left( J_1+M_1^
{\,\prime}-\frac{\omega _1}{2} \right)\left(J_2+M_2^{\,\prime}-\frac
{\omega _2}{2}\right)}
\lambda ^{2\left(z_1\left(J_2+M_2^{\,\prime}-
\frac{\omega _2}{2}\right)-z_2\left(J_1+M_1^{\,\prime}-\frac{\omega _1}
{2}\right)\right)}  \nonumber \\
&   &  \quad
\left\{ \sum _{n=0}^{\infty}\,\frac{\left(1-Q^{-2}\right)^n}{[n]!}\,
Q^{-\frac{1}{2}n(n-1)}\,Q^{n\left( J_1-J_2+M_1^{\,\prime}-M_2^{\,\prime}
-\frac{1}{2}\left(\omega _1 - \omega _2 \right)\right)}
\lambda ^{n
\left(z_1+z_2\right)} \right. \nonumber \\
&    &  \quad \left. \phantom{\sum _{n=0}^{\infty}}
\Pi _{l=1}^{n}\,\left[ \omega _1-J_1-M_1^{\,\prime}+l\right]
\left[J_2+M_2^{\,\prime}+l\right]
\delta _{M_1+n}^{M_1^{\,\prime}}
\delta _{M_2-n}^{M_2^{\,\prime}} \right\}\,.
\end{eqnarray}

\noindent It should be noted
that the presence of the second deformation parameter
$\lambda$ in (3.9) will produce many new $R$-matrices.  For nongeneric
values of $Q$, the matrix representation (3.9) acts as a source for
obtaining the finite dimensional two parametric$(p,q)$
nonstandard coloured $R$-matrices.
Let us consider the simplest example $\left( J_1=J_2=\frac{1}{2},
\right.$$\left. z_1=z_2=z \right)$ with different colour parameters in
the two sectors of the Hilbert space.  For $Q^2$ $=$ $-1$, we get

\begin{equation}
\bar{\cal R}^{\frac{1}{2}\,z\,\omega _1\,,\,\frac{1}{2}\,z\,\omega_2}
\,=\,c\left( \begin{array}{cccc}
             t_1 & 0 & 0 & 0 \\
             0 & s^{-1} & w & 0 \\
             0 & 0 & st_1t_2^{\,-1} & 0 \\
             0 & 0 & 0 & -t_2^{\,-1}
             \end{array} \right)\,,
\end{equation}

\noindent where $c = Q^{-\omega _2 \left( 1-\frac{\omega _1}{2} \right)}
\lambda ^{z \left( \omega _1-\omega _2 \right)}$\,,
$t_1 = -Q^{-\omega _1}$\,, $t_2 = -Q^{-\omega _2}$\,,
$s = \lambda ^{2z}$ and
$w = \left(t_1-t_1^{\,-1}\right)t_1^{\,\frac{1}{2}}t_2^{\,-\frac{1}{2}}$
\,.  This is an example of coloured $R$-matrix
and may be viewed as a generalization of the result obtained in [7].
When $\omega _1$ $=$ $\omega _2$ $=$ $\omega $  the $R$-matrix (3.10)
reduces to

\begin{equation}
\bar{\cal R}^{\frac{1}{2}\,z\,\omega ,\frac{1}{2}\,z\,\omega }\,\sim \,
\left( \begin{array}{cccc}
t & 0 & 0 & 0 \\
0 & s^{-1} & t-t^{-1} & 0 \\
0 & 0 & s & 0 \\
0 & 0 & 0 & -t^{-1}
\end{array} \right)\,.
\end{equation}

\noindent Apart from a scale factor, the matrix in (3.11) agrees with the
nonstandard $R$-matrix $\tilde{R}$ in (2.6) after the replacement $Q$
$\rightarrow$ $t$, $\lambda$ $\rightarrow$ $s$\,.  This completes
our discussion of the connection between the nonstandard two-parameter
$R$-matrix (2.1) with the universal ${\cal R}$-matrix of $U_{p,q}(gl(2))$.

\bigskip

\noindent {\bf 4. Superization of
$U_{p,q}\left( \zeta , H, X_\pm \right)$}

\medskip

\noindent Using a superization procedure [4] we now discuss the connection
between $U_{p,q}\left( \zeta ,H,X_\pm \right)$ and the super-Hopf algebra
$U_{p,q}(gl(1|1))$.  In [4] it is argued that if ${\cal H}$ is a Hopf
algebra containing a group-like element $g$ such that $g^2$ $=$ $1$,
then, there exists a super-Hopf algebra $\hat{\cal H}$ with identical
algebraic and counit structures while the coproduct, antipode and the
universal $R$-matrix of ${\cal H}$, $\left\{ \Delta (h) \left( = \sum
_{k}x_k \otimes y_k \right) \right.$, $S(h)$, ${\cal R} \left( = \sum_{k}
X_k \otimes Y_k \right) \mid $ $ \left. \left(h,x_k,y_k,X_k,Y_k\right) \in
{\cal H} \right\}$, are modified to the corresponding quantities of
$\hat{\cal H}$:

\renewcommand{\theequation}{4.{\arabic{equation}}}
\setcounter{equation}{0}

\begin{eqnarray}
\Delta \left( \hat{h} \right) & = & \sum_{k} x_k \otimes
g^{{\rm deg}\left( x_k \right)}y_k\,, \quad
S\left( \hat{h} \right)\,=\,S(h)g^{{\rm deg}(h)}\,, \nonumber \\
\hat{\cal R} & = & {\cal R}_g \sum_k X_k \otimes g^{{\rm deg}
\left( X_k \right)} Y_k\,,  \nonumber \\
&   &  \ \ {\cal R}_g\,=\,\frac{1}{2}( 1\!{\rm l} \otimes 1\!{\rm l} +
1\!{\rm l} \otimes g + g \otimes 1\!{\rm l} - g \otimes g )\,.
\end{eqnarray}

\noindent The map $\left\{ h \rightarrow \hat{h} \mid h \in {\cal H},
\hat{h} \in \hat{\cal H} \right\}$ preserves the algebraic structure and
it is understood that, in the right hand side of (4.1) the elements of
${\cal H}$, after simplification, are mapped to their hatted superpartners
in $\hat{\cal H}$.  The ${\rm deg}(h)$, $\forall h \in {\cal H}$, is
given by $ghg^{-1}$ $=$ ${\rm deg}(h)\,h$.  We exhibit these properties
in $(p,q)$-deformed examples.

The $R$-matrix

\begin{equation}
\hat{R}\,=\,\left(
\begin{array}{cccc}
Q^{-1} & 0 & 0 & 0 \\
0 & \lambda ^{-1} & 0 & 0 \\
0 & \sigma & \lambda & 0 \\
0 & 0 & 0 & Q
\end{array} \right)
\end{equation}

\noindent satisfies the super-YBE and is known [11] to describe the
`function' Hopf algebra $Fun_{p,q}(GL(1|1))$ (or $GL_{p,q}(1|1)$ ).
The universal ${\cal R}$-matrix
$\hat{\cal R}$ for the dually paired enveloping algebra $U_{p,q}(gl(1|1))$
[12] has been obtained by direct computation [13].  We can derive this
$\hat{\cal R}$ by applying Reshetikhin's technique [9] for introducing
multiple deformation parameters as follows.  The commutation relations
for the generators $\left( \hat{\zeta},\hat{H},\hat{X}_{\pm} \right)$ of
$U_{p,q}(gl(1|1))$ read

\begin{eqnarray}
\hat{X}_\pm ^{\,2} & = & 0\,, \quad
\left[ \hat{H} , \hat{X}_\pm \right] = \pm \hat{X}_\pm \,, \quad
\left\{ \hat{X}_+ , \hat{X}_- \right\} = \left[ 2 \hat{\zeta} \right]\,,
\nonumber \\
\left[ \hat{\zeta} , X \right] & = & 0\,,
\quad \forall X \in \left( \hat{H}, \hat{X}_\pm \right)\,,
\end{eqnarray}

\noindent and the coalgebraic structure is

\begin{eqnarray}
\Delta \left( \hat{X}_\pm \right) & = & \hat{X}_\pm \otimes
Q^{\hat{\zeta}}\lambda ^{\pm \hat{\zeta}} +
Q^{-\hat{\zeta}} \lambda ^{\mp \hat{\zeta}} \otimes \hat{X}_\pm \,,
\nonumber \\
\Delta \left( \hat{H} \right) & = & \hat{H} \otimes 1\!{\rm l} +
1\!{\rm l} \otimes \hat{H}\,, \qquad
\Delta \left( \hat{\zeta} \right)\,=\,\hat{\zeta} \otimes
1\!{\rm l} + 1\!{\rm l} \otimes \hat{\zeta}\,, \nonumber \\
\epsilon (X) & = & 0\,, \qquad  S(X)\,=\,-X\,, \quad \forall X \in
\left( \hat{\zeta}, \hat{H}, \hat{X}_\pm \right)\,.
\end{eqnarray}

\noindent The coproduct relations (4.4) reveal that a structure
similar to (2.29) holds with the choice

\begin{equation}
\hat{F}\,=\,\lambda ^{\left( \hat{H} \otimes \hat{\zeta} -
\hat{\zeta} \otimes \hat{H} \right)}\,.
\end{equation}

\noindent Now, using the known universal ${\cal R}$-matrix of
$U_Q(gl(1|1))$ [4]

\begin{equation}
\hat{\cal R}_{\lambda =1}\,=\,Q^{2\left( \hat{\zeta} \otimes \hat{H}
+ \hat{H} \otimes \hat{\zeta} \right)} \left( 1\!{\rm l} \otimes
1\!{\rm l} + \sigma Q^{\hat{\zeta}} \hat{X}_+ \otimes Q^{-\hat{\zeta}}
\hat{X}_- \right)
\end{equation}

\noindent and the prescription (2.31), we obtain the universal
${\cal R}$-matrix for $U_{p,q}(gl(1|1))$ as

\begin{equation}
\hat{\cal R} = Q^{2 \left( \hat{\zeta} \otimes \hat{H} + \hat{H}
\otimes \hat{\zeta} \right)} \lambda ^{2 \left( \hat{\zeta} \otimes
\hat{H} - \hat{H} \otimes \hat{\zeta} \right)} \left( 1\!{\rm l} \otimes
1\!{\rm l} + \sigma Q^{\hat{\zeta}} \lambda ^{\hat{\zeta}} \hat{X}_+
\otimes Q^{-\hat{\zeta}} \lambda ^{\hat{\zeta}} \hat{X}_- \right)
\end{equation}

\noindent which satisfies the super-YBE with the composition rule for
the graded operators $(A,B,C,D)$ $\in $ $U_{p,q}(gl(1|1))$

\begin{equation}
(A \otimes B)(C \otimes D)\,=\,(-1)^{{\rm deg}(B){\rm deg}(C)}
(AC \otimes BD)\,.
\end{equation}

\noindent Our expression for $\hat{\cal R}$ in (4.7) is seen to agree
with the result obtained in [13] by direct computation.

Comparing (2.18) and
(2.22) with (4.3) and (4.4), while introducing the map
$\left( \zeta , H, X_\pm
\right)$ $\rightarrow $ $\left( \hat{\zeta}, \hat{H}, \hat{X}_\pm
\right)$, it follows that the commutation relations and the counit
maps for the super-Hopf algebra
$U_{p,q}(gl(1|1))$ are identical to those of $U_{p,q}\left( \zeta ,
H, X_\pm \right)$.  The coproduct (4.4), antipode (4.4)
and the universal ${\cal R}$-matrix (4.7) for
$U_{p,q}(gl(1|1))$ follow from the analogous formulae, (2.20), (2.23)
and (2.32), for  $U_{p,q}\left( \hat{\zeta}, \hat{H}, \hat{X}_\pm \right)$
according to the prescription (4.1).
The quantity $(-1)^{(\zeta - H) \otimes (\zeta - H)}{\cal R}_g$ is
central in nature as may be observed by direct computation.  It may,
therefore, be dropped while mapping ${\cal R}$ to $\hat{\cal R}$.

\bigskip

\noindent {\bf 5. A $(p,q)$-deformed super-Heisenberg algebra}

\medskip

\noindent Finally, we use the contraction technique \`{a} la Celeghini
et al. [18,19] to extract a two-parametric deformed super-Heisenberg
algebra $U_{p,q}(sh(1))$ as a limiting case of $U_{p,q}(gl(1|1))$.
To this end, let us  scale the generators and the quantization parameters
as

\renewcommand{\theequation}{5.{\arabic{equation}}}
\setcounter{equation}{0}

\begin{equation}
\hat{H}\,=\,\frac{1}{2\varepsilon} {\sf h} - N\,, \quad
\hat{\zeta}\,=\,\frac{\xi}{2\varepsilon}\,, \quad
\hat{X}_\mp \,=\,\frac{1}{\sqrt{\varepsilon}}\,c_\pm \,,\quad
Q\,=\,{\rm e}^{\varepsilon \Omega}\,, \quad
\lambda \,=\,{\rm e}^{\varepsilon \nu}\,.
\end{equation}

\noindent The commutation relations and the coproduct rules for
$U_{p,q}(sh(1))$, with $\left\{ \xi ,{\sf h}, N, c_\pm \right\}$ as the
generators, are obtained by studying the $\varepsilon$
$\rightarrow$ $0$ limit of the correponding structures of
the algebra $U_{p,q}(gl(1|1))$ in (4.3) and (4.4) respectively.
The result is

\begin{eqnarray}
c_\pm ^{\,2} & = & 0\,, \qquad \left[ N , c_\pm \right]\,=\,
\pm c_\pm \,, \qquad \left\{ c_+ , c_- \right\}\,=\,\Omega ^{-1}
{\rm sinh}\,\Omega \xi \,, \nonumber \\
\left[\,{\sf h} , X\,\right] & = & 0\,, \qquad [\xi , X\,]\,=\,0\,,
\quad \forall X \in U_{p,q}(sh(1))\,,
\end{eqnarray}

\noindent and

\begin{eqnarray}
\Delta \left( c_\pm \right) & = & c_\pm \otimes {\rm e}^
{\frac{1}{2} \left( \Omega \mp \nu \right) \xi } + {\rm e}^
{-\frac{1}{2} \left( \Omega \mp \nu \right) \xi } \otimes c_\pm \,,
\nonumber \\
\Delta (\xi ) & = & \xi \otimes 1\!{\rm l} + 1\!{\rm l}
\otimes \xi \,, \quad
\Delta ({\sf h})\,=\,{\sf h} \otimes 1\!{\rm l}
+ 1\!{\rm l} \otimes {\sf h}\,,  \nonumber \\
\Delta (N) & = & N \otimes 1\!{\rm l} + 1\!{\rm l} \otimes N\,.
\end{eqnarray}

\noindent In the contraction limit, the universal ${\cal R}$-matrix
(4.7) of $U_{p,q}(gl(1|1))$ yields, after the removal of a constant
singular factor, the universal ${\cal R}$-matrix of $U_{p,q}(sh(1))$:

\begin{equation}
{\cal R}^{sh}\,=\,{\rm e}^{\nu (N\otimes \xi - \xi \otimes N) -
\Omega (N \otimes \xi + \xi \otimes N)} \left( 1\!{\rm l} \otimes
1\!{\rm l} - 2\Omega {\rm e}^{\frac{1}{2}(\Omega + \nu )\xi } c_-
\otimes {\rm e}^{-\frac{1}{2}(\Omega - \nu )\xi } c_+ \right)\,.
\end{equation}

A spectral parameter dependent
${\cal R}$-matrix may be obtained from the universal ${\cal R}$-matrix
(5.4).  To this end, following [18], we define the map

\begin{equation}
T_x c_\pm \,=\,x^{\mp 1}c_\pm \,, \qquad
T_x{\sf h}\,=\,{\sf h}\,, \qquad
T_x \xi \,=\,\xi \,, \qquad
T_xN\,=\,N\,,
\end{equation}

\noindent and let

\begin{eqnarray}
{\cal R}^{sh}(x) & = & \left( T_x \otimes 1\!{\rm l} \right)
{\cal R}^{sh} \nonumber \\
& = & {\rm e}^{\nu (N \otimes \xi - \xi \otimes N) - \Omega
(N \otimes \xi + \xi \otimes N)} \nonumber \\
&    &  \ \  \times  \left( 1\!{\rm l} \otimes
1\!{\rm l} - 2 \Omega x\,{\rm e}^{\frac{1}{2} ( \Omega +
\nu )\xi }c_- \otimes {\rm e}^{-\frac{1}{2}( \Omega - \nu )\xi}
c_+ \right)\,.
\end{eqnarray}

\noindent A direct calculation then proves that the matrix
${\cal R}^{sh}(x)$ satisfies the spectral parameter dependent YBE

\begin{equation}
{\cal R}^{sh}_{12}(x){\cal R}^{sh}_{13}(xy){\cal R}^{sh}_{23}(y)\,
=\,{\cal R}^{sh}_{23}(y){\cal R}^{sh}_{13}(xy){\cal R}^{sh}_{12}(x)\,.
\end{equation}

\bigskip

\noindent {\bf 6. Conclusion}

\medskip

\noindent To conclude, we have studied the dually paired Hopf algebras
$A_{p,q}(R)$ and $U_{p,q}\left( \zeta ,H,X_\pm \right)$ associated
with the nonstandard $R$-matrix (2.1) involving two independent
parameters $p$ and $q$.  The universal ${\cal R}$-matrix of
$U_{p,q}\left( \zeta , H,X_\pm \right)$ has been obtained.
We have demonstrated an explicit construction of the nonstandard
$R$-matrix through a coloured generalized boson representation of the
universal ${\cal R}$-matrix of $U_{p,q}(gl(2))$ corresponding to the
nongeneric case $pq$ $=$ $Q^2$ $=$ $-1$.  In this example, by choosing
different colour parameters for the two sectors of nongeneric
representations of $U_{p,q}(gl(2))$ with a dimension 2, we have obtained
a two-parametric coloured $R$-matrix (3.10), which satisfies a coloured
YBE.  More importantly, the finite
dimensional representation (3.9) of the ${\cal R}$-matrix gives a
recipe for obtaining nonstandard two-parametric coloured $R$-matrices
for nongeneric values of $Q$.  Superization process describes a map
between $U_{p,q}\left( \zeta ,H,X_\pm \right)$ and $U_{p,q}(gl(1|1))$.
A contraction procedure has been used to obtain a $(p,q)$-deformed
quasitriangular super-Heisenberg algebra $U_{p,q}(sh(1))$.

\vspace{1cm}

\baselineskip12pt

{\bf References}

\begin{enumerate}

\item V.G. Drinfeld: Proc. ICM-86 (Berkeley) 1987;
M. Jimbo: Lett. Math. Phys. 10 (1985) 63

\item H.C. Lee, M. Couture, N.C. Schmeing: `Connected link
polynomials', Chalk-river preprint 1988;
M.L. Ge, L.Y. Wang, K. Xue, Y.S. Wu: Int. J. Mod. Phys.
A 4 (1989) 3351

\item N. Jing, M.L. Ge, Y.S. Wu: Lett. Math. Phys. 21 (1991) 193

\item S. Majid, M.J. Rodriguez-Plaza: `Universal ${\cal R}$-matrix
for a nonstandard quantum group and superization', Preprint DAMTP
91/47

\item S. Majid: `Transmutation theory and rank for quantum braided
groups' Preprint DAMTP 91/10

\item A.J. Macfarlane: J. Phys. A: Math. Gen. 22 (1989) 4551;
L.C. Biedenharn: J. Phys. A: Math. Gen. 22 (1989) L873;
C.P. Sun, H.C. Fu: J. Phys. A: Math. Gen. 22 (1989) L983;
T. Hayashi: Commun. Math. Phys. 127 (1990) 129

\item M.L. Ge, C.P. Sun, K. Xue: Int. J. Mod. Phys. A 7 (1992) 6609

\item P.P. Kulish: Zap. Nauch. Semin. LOMI 180 (1990) 89;
E.E. Demidov, Yu.I. Manin, E.E. Mukhin, D.V. Zhadanovich: Prog.
Theor. Phys. Suppl. 102 (1990) 203;
A. Sudbery: J. Phys. A: Math. Gen. 23 (1990) L697;
M. Takeuchi: Proc. Japan Acad. 66 (1990) 112

\item N.Yu. Reshetikhin: Lett. Math. Phys. 20 (1990) 331

\item A. Schirrmacher, J. Wess, B. Zumino: Z. Phys. C: Particles and
Fields 49 (1991) 317;
A. Schirrmacher: J. Phys. A: Math. Gen. 24 (1991) L1249;
C. Burdick, L. Hlavaty: J. Phys. A: Math. Gen. 24 (1991) L165

\item R. Chakrabarti, R. Jagannathan: J. Phys. A: Math. Gen. 24 (1991)
5683

\item L. Dobrowski, L.Y. Wang: Phys. Lett. B226 (1991) 51

\item C. Burdick, P. Hellinger: J. Phys. A: Math. Gen. 25 (1992) L1023

\item V.K. Dobrev: J. Geom. Phys. 11 (1993) 367;
C. Fronsdal, A. Galindo: Lett. Math. Phys. 27 (1993) 59;
R. Barbier, J. Meyer, M. Kibler: J. Phys. G: Nucl. Part. 20 (1994) L13

\item R. Chakrabarti, R. Jagannathan: J. Phys. A: Math. Gen. 27 (1994)
2023

\item M. Bednar, C. Burdick, M. Couture, L. Hlavaty: J. Phys. A: Math.
Gen. 25 (1992) L341

\item N.Yu. Reshetikhin, L.A. Takhtajan, L.D. Faddeev: Leningrad Math.
J. 1 (1990) 193

\item E. Celeghini, R. Giachetti, E. Sorace, M. Tarlini: J. Math. Phys.
31 (1991) 2548

\item E. Celeghini, R. Giachetti, E. Sorace, M. Tarlini: J. Math. Phys.
32 (1992) 1115

\end{enumerate}

\noindent
{\em Note added}:  We are thankful to Prof. V.K. Dobrev for bringing
our attention to the papers: (1) V.K. Dobrev, "Introduction to
Quantum Groups", G\"{o}ttingen University preprint, (April 1991),
to appear in: {\it Proc. 22nd Ann. Iranian Math. Conf.} (March 1991,
Mashhad) and (2) V.K. Dobrev, {\it J. Math. Phys. 33 (1992)
3419}, where the structure of $U_{p,q}(gl(2))$ dual to $GL_{p,q}(2)$
was established earlier by him (see Refs. in [14]) using an approach
independent of Schirrmacher, Wess and Zumino [10]. It may also be noted
that the work of Fronsdal and Galindo (see Refs. in [14]) on the
$U_{p,q}(gl(2)) \longrightarrow GL_{p,q}(2)$ exponential relationship
is based on a different approach.

\end{document}